\documentclass[11pt,a4paper]{article}
\usepackage{a4wide}

\usepackage{amsmath,amssymb}
\usepackage{mathtools}
\usepackage{siunitx}
\usepackage{bm}
\usepackage{booktabs}
\usepackage{slashbox}

\DeclareMathOperator*{\minimize}{minimize}
\newcolumntype{C}[1]{>{\centering\arraybackslash}p{#1}}

\usepackage[dvipdfmx]{graphicx}

\usepackage{color}

\title{%
Koopman-Model Predictive Control with Signal Temporal Logic Specifications for Temperature Regulation of a Warm-Water Supply System\thanks{Corresponding author: Yoshihiko Susuki ({\tt susuki.yoshihiko.5c@kyoto-u.ac.jp})}
}
\author{%
Ryo~Miyashita\footnote{Department of Electrical and Information Systems, Osaka Prefecture University, Sakai, Japan}, 
Yoshihiko~Susuki\footnote{Department of Electrical Engineering, Kyoto University, Kyoto, Japan} 
and Atsushi~Ishigame\footnote{Department of Electrical and Electronic Systems Engineering, Osaka Metropolitan University, Sakai, Japan}
}
\date{}

\begin{document}
\maketitle

\begin{abstract}
Control of warm-water supply for dialysis treatment in a hospital environment is typical of safety-critical control problems. 
In order to guarantee the continuity of warm-water supply satisfying physical specifications for a wide range of operating conditions, it is inevitable to consider the nonlinearity involved in a dynamic model of a warm-water supply system for the control design. 
In this paper, we propose to incorporate control specifications described by signal temporal logic, which is a temporal logic with semantics over finite-time signals in formal methods, into the so-called Koopman-Model Predictive Control (MPC) as a novel technique of nonlinear MPC based on the Koopman operator framework for nonlinear systems. 
This enables us to generate a sequence of optimal inputs such that the controlled state of a nonlinear system can satisfy the specifications.
The proposal is applied to the temperature regulation of warm-water supply, and its effectiveness is established numerically.
\end{abstract}

\section{Introduction}
\label{s:1}

Design of energy control systems in medical institutions is typical of safety-critical control problems in the modern aging society. 
In \cite{SCCTA}, the second author of the present paper introduced a problem of control for warm-water supply in a safety-critical hospital environment. 
A system model of warm-water supply based on the practical hospital for dialysis treatment in Japan was addressed.
In the present paper, we focus on the same control problem and report a new control design by a combination of techniques from nonlinear Model Predictive Control (MPC) and formal synthesis.

The formal synthesis, in which the goal is to synthesize or control a finite-state transition system from a temporal logic specification, has attracted interest with connection to control of dynamical systems: see, e.g., \cite{Belta,Ushio}. 
Temporal logics have been widely utilized for specifying desired behaviors of the transition system \cite{Belta}. 
Temporal logics with semantics over finite-time signals, such as metric temporal logic \cite{MTL_original}
and Signal Temporal Logic (STL) \cite{STL2004,STL2010}, have been connected to the so-called optimal control of dynamical systems such as MPC with STL specifications for plants with LTI dynamics \cite{MPCSTL,BluSTL}. 

In the preliminary report \cite{miyashita},
we focused on the optimization-based method \cite{MPCSTL,BluSTL} for the control of warm-water supply dynamics. 
The dynamics were originally modeled by a nonlinear dynamic model \cite{Otani}, and thus an LTI model was derived through linearization around a nominal operating point (that is, stable equilibrium point of that model) and used for the control design. 
The method was effective for certain initial conditions, but it showed limitations in a situation where the state of the nonlinear model is far from the nominal operating point. 
Precisely speaking, for this situation the optimization does not often converge. 
It is thus necessary to perform the control design for a wider range of operating conditions in order to guarantee the safety of warm-water supply.

In the present paper, we report a new control design of the warm-water supply system by considering the nonlinearity of the dynamic model.
The main contributions of this paper are two-fold.
First, to consider the nonlinearity for the design, we propose to incorporate control specifications described by STL into the so-called Koopman-MPC \cite{KMPC}. 
The Koopman-MPC is a novel technique of nonlinear MPC based on the Koopman operator framework \cite{Sbook} for nonlinear dynamical systems and is used in various applications such as power grid \cite{Powergrid}. 
The main idea of Koopman-MPC is to utilize a linear model (called linear predictor)  that predicts the time evolution of the state of a target nonlinear system. The linear predictor is mathematically guided by the action of the Koopman operator defined for the target nonlinear system and can be estimated directly from time-series data on the state as well as input (see Sections 3.2 and 4.2), which are termed as Koopman linearization. 
The incorporation of STL specifications into Koopman-MPC is novel to the best of the authors' survey.
Second, we numerically show its effectiveness for the control design of warm-water supply by applying it to the nonlinear model of the system. 
A procedure of the control design is presented, for example, the choice of observable functions to derive the linear predictor.
This paper is a refined version of the non-reviewed proceeding \cite{
miyashita2}. 

The structure of this paper is as follows. 
Section~\ref{s:2} describes the warm-water supply system and its nonlinear dynamic model.  
Section~\ref{s:3} describes the Koopman-MPC with STL specifications which we propose in this paper. 
Section~\ref{s:4} applies the proposed method 
to the nonlinear model of the warm-water supply system. 
Section~\ref{s:5} is the conclusion of this paper with a brief summary and future directions.

\section{Model of warm-water supply system}
\label{s:2}

This section provides an overview of the warm-water supply system considered in this paper and its nonlinear dynamic model representing global dynamics of the system.
\begin{figure}[t]
\begin{center}
\includegraphics[width=8cm]{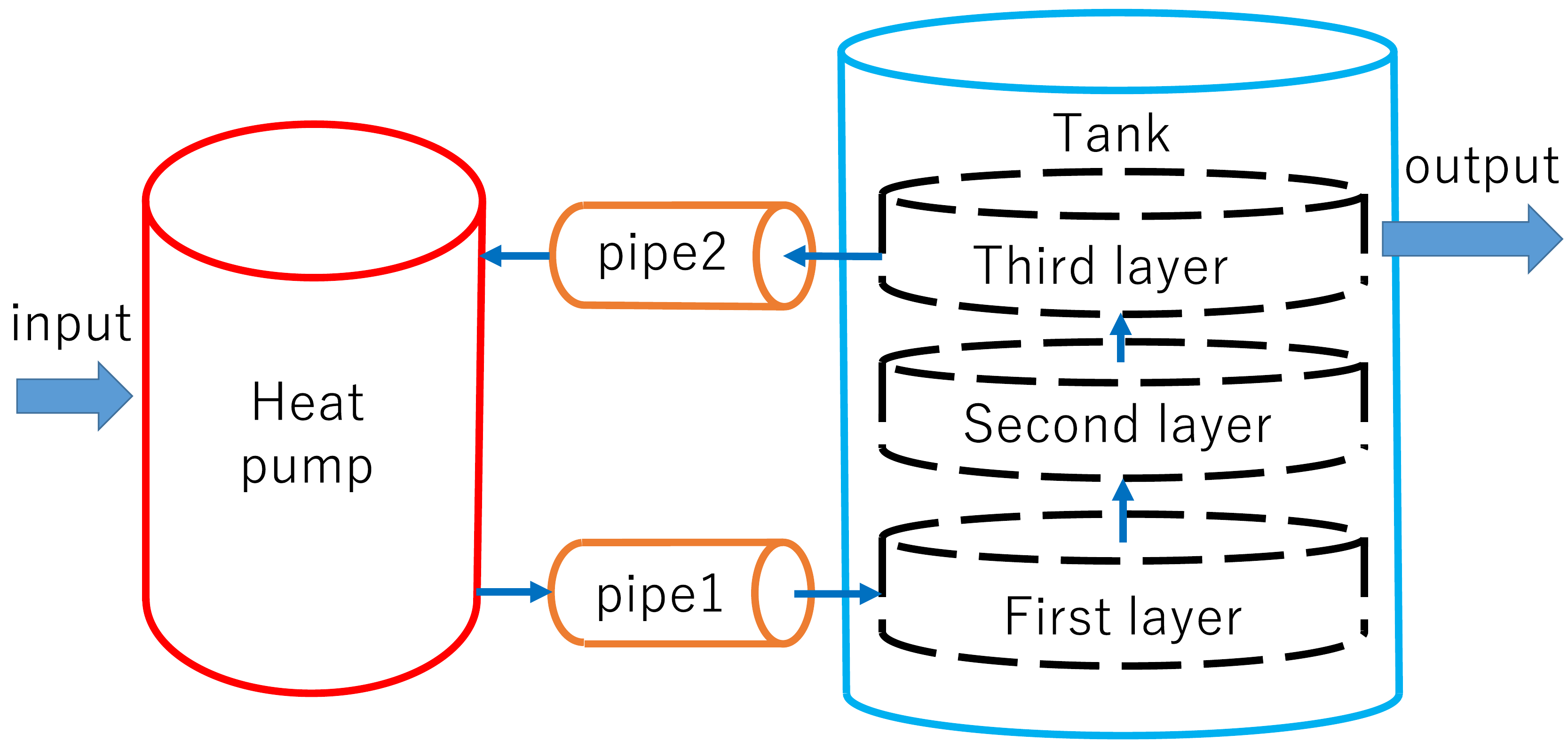} 
\caption{Structure of warm-water supply system consisting of one heat pump, two pipes, and one tank}
\label{f:loop}
\end{center}
\end{figure}

\subsection{System model and control specifications}
\label{s:2.1}

Figure~\ref{f:loop} shows the warm-water supply system which we consider in this paper. 
The system is related to the actual medical institution \cite{matsuo} and consists of one heat pump, two pipes, and one tank. 
The heat pump aims to raise the temperature of water. 
The pipes flow warm-water between the heat pump and the tank. 
The tank stores warm-water and supplies it to the load assumed to be connected to the upper part of the tank. 
The load is supposed to be ideal, and its characteristics can be neglected. 
Also, to consider temperature gradient of water inside the tank, we use the three-layer model in \cite{yokoyama} and describe it in Figure~\ref{f:loop}. 
Between nearest layers, the exchange of heat is considered. 

In this paper, we introduce two control specifications based on practical constraints of the system.
Dialysis treatment requires warm water of about the temperature of human skin (about $37\,\si{\degreeCelsius}$) \cite{matsuo}. 
Based on this, the first control specification is to keep the temperature of the third layer of the tank, which will be regarded as the output of a nonlinear dynamic model, above $40\,\si{\degreeCelsius}$. 
The value is set by considering the temperature drop from the third layer to the (ideal) load. 
The second control specification is based on the operating range of the heat pump as described in \cite{irie}, where its electricity consumption ranges from $80\%$ of the rated to the rated itself. 
The rated is set to $26.5\,\si{kW}$ based on actual measurement data of the heat pump in the hospital. 

\subsection{Nonlinear dynamic model}
\label{s:2.2}
The dynamic characteristics of the warm-water supply system are modeled in \cite{Otani} based on \cite{yokoyama,irie}. 
The model is represented by the nonlinear dynamical system of Eq.\,(\ref{nonmodel}):
\begin{equation}
\label{nonmodel}
\left.
\begin{aligned}
\dot{\bm{x}} &= \bm{f} \left (\bm{x},u,w \right) \\
y &= g(x)
\end{aligned}
\right\}.
\end{equation}
For the six-dimensional state vector $\bm{x}$, the elements $x_1,\ldots,x_6$ are the water temperatures of every component in the following order: heat pump, pipe 1, first layer to third layer of the tank, and pipe 2. 
The nonlinear vector-valued function $\bm{f}$ on the right-hand side of Eq.\,(\ref{nonmodel}) is given in Appendix~A. 
The output function $g$ (thus the output variable $y$) corresponds to $x_5$ since the output is the water temperature of the third layer of tank. 
The input variable $u$ represents the power input to the heat pump, and the disturbance $w$ does the temperature of environment surrounding the system.
The variables in Eq.\,(\ref{nonmodel}) and their physical meanings are summarized in Table~\ref{tb:EoSvariable}. 
It should be mentioned that the nonlinearities in Eq.\,(\ref{nonmodel}) are the squared and cubic terms related to several elements of the state vector. 
Eq.\,(\ref{nonmodel}) is used to simulate the plant and estimate the linear predictor in Section~\ref{s:4}.

\begin{table}[t]
\caption{Variables of the nonlinear dynamic model}
\label{tb:EoSvariable}
\begin{center}     
\begin{tabular}{cccl} \toprule[0.5mm]
&  variable & dimension & physical meaning  \\ \hline 
$\bm{x}$   &state & 6 &
temperature of water 
in the components \\ \hline
$u$ & input & 1  &
power input to 
heat pump \\ \hline 
$w$    &disturbance&1 &outside temperature\\ \hline
$y$    & output & 1 & 
temperature of water 
of the 3rd layer of tank \\\bottomrule[0.5mm]
\end{tabular}
\end{center}
\end{table}

\section{Koopman-MPC with STL specifications}
\label{s:3}

This section is devoted to the first contribution of this paper---to propose the Koopman-MPC with STL specifications. 
First of all, we briefly explain the STL and Koopman-MPC, and then describe our idea of how to incorporate a control specification described by STL into the formulation of Koopman-MPC. 

\subsection{STL (Signal Temporal Logic)}
\label{s:3.1}

STL is a temporal logic that allows the specification of properties of  real-valued signals and the automatic generation of monitors for testing these properties 
\cite{STL2004,STL2010}. 
STL formulas are defined recursively in \cite{MPCSTL,BluSTL} as follows:
\begin{equation}
\phi ::= 
\mu~|~\lnot\mu~|~\phi_1\land\phi_2~|~\phi_1\lor\phi_2~|~\square_{[a,b]}\phi~|~\phi_1{\cal U}_{[a,b]}\phi_2, 
\end{equation}
where $\mu$ is a predicate whose value is determined by the sign of a function of an underlying signal, and $\phi$ is an STL formula. 
We here use the usual logical operators such as 
$\lnot$(negation), $\land$(conjunction), $\lor$(disjunction), and $\to$(implication). 
The temporal logic operators are $\square_{[a,b]}\phi$ (always: if the property defined by $\phi$ holds (that is, $\phi$ is true ($\top$)) at any time between $a$ and $b$) and $\phi_1{\cal U}_{[a,b]}\phi_2$ (until: if $\phi_1$ holds at any time before $\phi_2$ holds, and $\phi_2$ holds at some onset between $a$ and $b$). 
Additionally, the operator 
$\Diamond_{[a,b]}\phi:=\top{\cal U}_{[a,b]}\phi$ is defined (eventually: if $\phi$ holds at some onset from $a$ to $b$). 
For details including its semantics and boundedness property, see \cite{MPCSTL,BluSTL}.

\subsection{Koopman-MPC}
\label{s:3.2}

The Koopman-MPC is the MPC for nonlinear dynamical systems 
using linear predictors based on the so-called Koopman linearization \cite{KMPC}. 
A main idea of constructing an accurate linear predictor is to introduce a mapping from the state space of a target nonlinear system to a space with dimension sufficiently higher than that of the original system.  
To make this concrete, a function: $\mathbb{X} \rightarrow \mathbb{C}$ is introduced from the state space $\mathbb{X}$ of the nonlinear system \eqref{nonmodel} to $\mathbb{C}$, called the {\textit{observable}}. 
A linear predictor is estimated using a set of observables, $\psi_i:i = 1,\ldots,N$, where $N$ is sufficiently larger than the dimension of $\mathbb{X}$.  
This estimation can be performed based on time-series data of the state $\bm{x}_k=\bm{x}(kh)$, input $u_k=u(kh)$, and disturbance $w_k=w(kh)$ for \eqref{nonmodel} under the sampling period $h$, and is explained in Section~\ref{s:4.2}. 
An advantage of the Koopman linearization is that it can work globally in the state space $\mathbb{X}$, unlike the conventional local linearization around a particular state such as an equilibrium point.  
The linear predictor \eqref{K_LTI} for \eqref{nonmodel} is derived in discrete-time form:
\begin{equation}
\label{K_LTI}
\left.
\begin{array}{rcl}
\bm{z}_{k+1}&=&{\mathsf{A}}\bm{z}_k+\bm{b}_{u}u_k+\bm{b}_{d}w_k \\
\hat{\bm{x}}_k&=&{\mathsf{C}}\bm{z}_k
\end{array}
\right\},
\end{equation}
where $k$ is the discrete time. 
The state $\bm{z}\in{\mathbb{C}}^N$ is defined by the set of observables $\psi$ as $\bm{z}\coloneqq \bm{\psi}(\bm{x})= \left[\psi_1(\bm{x}),\ldots,\psi_N(\bm{x})\right]^{\rm T}$  ($\rm T$ stands for the transpose operation). 
The output $\hat{\bm{x}}$ is the predicted value of the sampled $\bm{x}$. 
The coefficient matrices ${\sf A},\bm{b}_u,\bm{b}_d,{\sf C}$ are determined through the data-based estimation.

The Koopman-MPC controller then solves at each time step $k$ of the closed-loop operation the following optimization problem:
\begin{equation}
\label{KoopMPC}
\left.
\begin{array}{cc}
\displaystyle{\minimize_{u_0,\ldots,u_{N_{\rm{p}}-1}}}
&{\it{J}}\left( \left( u_{i} \right )_{i=0}^{N_{\rm{p}}-1},\left( \bm{z}_{i} \right)_{i=0}^{N_{\rm{p}}-1}\right) \\
\text{subject to} &\bm{z}_{i+1} = {\mathsf{A}}\bm{z}_{i} + \bm{b}_{u}u_{i} + \bm{b}_{d}w_{i} \\
&u_{\text{min}} \leq u_i \leq u_{\text{max}} \\
&\bm{z}_{\text{min}} \leq \bm{z}_i \leq \bm{z}_{\text{max}} \\
&  (i =0,\ldots,N_{\rm{p}}-1) \\
\text{parameter}& \bm{z}_0=\bm{\psi}(\bm{x}_{k})
\end{array}
\right\},
\end{equation}
where $N_{\rm{p}}$ is the number of prediction steps (horizon), and $J$ is a cost function. 
For an optimizer at time step $k$, denoted by $u^\star_0,u^\star_1,\ldots,u^\star_{N_{\rm{p}-1}}$, the initial control $u^\star_0$ is used at the time step $k$. 
Note that because \eqref{nonmodel} is in continuous-time form, we apply the above input to \eqref{nonmodel} using the zeroth-order hold.

\subsection{Proposed method: Koopman-MPC with STL specifications}
\label{s:3.3}

In this paper, we propose to incorporate an STL specification with the Koopman-MPC. 
The controller solves the following optimization problem: 
\begin{equation}
\label{KSTLMPC}
\left.
\begin{array}{cc}
\displaystyle{\minimize_{u_0,\ldots,u_{N_{\rm p}-1}}} & {\it{J}}\left( \left( u_{i} \right )_{i=0}^{N_{\rm{p}}-1},\left( \bm{z}_{i} \right)_{i=0}^{N_{\rm{p}}-1}\right) \\
\text{subject to}& \bm{z}_{i+1} = {\mathsf{A}}\bm{z}_{i} + \bm{b}_{u}u_{i} + \bm{b}_{d}w_{i} \\
&u_{\text{min}} \leq u_i \leq u_{\text{max}} \\
&\bm{z}_{\text{min}} \leq \bm{z}_i \leq \bm{z}_{\text{max}} \\
&(i=0,\ldots,N_{\rm{p}}-1)\\
&\bm{z}_{0},u_{0}\vDash \bm{\phi} \\
\text{parameter}& \bm{z}_0=\bm{\psi}(\bm{x}_{k})
\end{array}
\right\},
\end{equation}
where $\bm{\phi}$ stands for STL formulas corresponding to control specifications. 
The key inclusion $\bm{z}_0,u_{0}\vDash \bm{\phi}$ implies that $\bm{z}_{0},u_{0}$ satisfy the STL formulas $\bm{\phi}$ at time step $k$ of the closed-loop operation. 
The problem \eqref{KSTLMPC} except for the parameter linked to the original nonlinear system is basically same as that for the MPC with STL specifications for discrete-time LTI systems \cite{MPCSTL,BluSTL}. 
In these, the system dynamics as well as the STL formulas are automatically encoded to Mixed-Integer Linear Program (MILP) constraints. 
The MILP encoding is available from the MATLAB toolbox {\tt BluSTL} \cite{BluSTL}, by which the Koopman-MPC with STL specifications can be implemented.

\section{Application to warm-water supply system}
\label{s:4}

This section is devoted to the second contribution of this paper---to apply the Koopman-MPC with STL specifications to the control problem of the warm-water supply system.

\subsection{Description of control specifications with STL}
\label{s:4.1}

We describe the control specifications described in Section~\ref{s:2.1} as STL formulas. 
For simplicity of the presentation, all units in the STL formulas are omitted. 
The first specification of the water temperature $y_k=y(kh)$, denoted as $\phi_1$, is given by 
\begin{equation}
\label{stltank3}
\phi_1 = \square_{[420,{\rm{end}}]} \left( y_k \geqq 40 \right), 
\end{equation}
where `end' stands for the finite, final time of closed-loop operation. 
Eq.\,\eqref{stltank3} guarantees the supply of warm water with temperature larger than and equal to $40\,\si{\degreeCelsius}$. 
The prescribed time in the formula, $420\,\si{s}$, is set by taking into account a transient duration for the water temperature to rise. 
The second specification $\phi_2$ for the power input to the heat pump is given by
\begin{equation}
\label{stlhp}
\phi_2 = \square_{[0,{\rm{end}}]}
\bigl[ \left \{ (u_k > 0.001) \land (u_k < 0.01) \right \} 
\lor  \left \{ (u_k \geqq 21.2) \land (u_k \leqq 26.5) \right \} \bigr]. 
\end{equation}
At any time, the heat pump will consume $0.001\,\si{kW}$ to $0.01\,\si{kW}$, that is, $0\,\si{kW}$ of power when it stops, and it will also consume $21.2\,\si{kW}$ to $26.5\,\si{kW}$ of power when it runs. 

\subsection{Setting of Koopman-MPC}
\label{s:4.2}

Here, we explain the estimation of linear predictor for the warm-water supply system.
For this, the following set of observables $\bm{\psi}$ is introduced in which all nonlinear terms of the state in the original nonlinear model \eqref{nonmodel} are taken out, referring to \cite{Marcos}:
\begin{equation}
\label{lifting}
\bm{\psi}(\bm{x}) = [ x_1,x_2,x_3,x_4,x_5,x_6, 
x_3^2,x_4^2,x_5^2,x_3^2x_4,x_3x_4^2,x_4^2x_5,x_4x_5^2, x_3^3,x_4^3,x_5^3  ]^{\rm T}.
\end{equation}
The estimation procedure is based on \cite{KMPC} and summarized below. 
The initial dataset \eqref{dataset} with number $K=10000$ is introduced as 
\begin{equation}
\label{dataset}
\left.
\begin{aligned}
\mathsf{X} &:= [\bm{x}^{(1)}_0, \bm{x}^{(2)}_0, \ldots, \bm{x}^{(K)}_0] \\
\bm{u} &:= [u^{(1)}_0, u^{(2)}_0, \ldots, u^{(K)}_0] \\
\bm{w} &:= [1,1,\ldots,1]w_0
\end{aligned}
\right\},
\end{equation}
where every element of $\bm{x}^{(i)}_0$ is commonly set at a uniform random number in $[10,40]$, and $u^{(i)}_0$ is set at either a uniform random number in $[21.2,26.5]$ or $0$. 
The probability of appearance of $0$ is $20\%$.
The disturbance is common for all the initial conditions, and $w_0$ is set at $10$. 
For each $\bm{x}^{(i)}_0$, its one-step ($h$) progress is computed using $u^{(i)}_0$ and $w_0$, denoted as $\bm{x}^{(i)}_1$, and the following dataset is additionally introduced:
\begin{equation}
{\sf X}' := [\bm{x}^{(1)}_1,\bm{x}^{(2)}_1,\ldots,
\bm{x}^{(K)}_1].
\end{equation}
The coefficient matrices of the linear predictor are then estimated with 
\begin{equation}
\label{solABC}
\left.
\begin{aligned}
[{\mathsf{A}},\bm{b}_u,\bm{b}_d] &= \mathsf{X}'_{\rm{lift}}
\begin{bmatrix}
\mathsf{X}_{\rm{lift}} \\
\bm{u} \\
\bm{w}
\end{bmatrix}^{\dag} \\
\mathsf{C}& = \mathsf{X}\mathsf{X}^{\dag}_{\rm{lift}}
\end{aligned}
\right\},
\end{equation}
where $\dag$ is the pseudo-inverse of a matrix, and $\mathsf{X}_{\rm{lift}}$ and $\mathsf{X}'_{\rm{lift}}$ are defined through $\bm{\psi}$ as 
\[
\label{lift}
\left.
\begin{aligned}
\mathsf{X}_{\rm{lift}} &= \bm{\psi}({\sf X})
:=[\bm{\psi}(\bm{x}^{(1)}_0), \bm{\psi}(\bm{x}^{(2)}_0), \ldots, \bm{\psi}(\bm{x}^{(K)}_0)] \\
\mathsf{X}'_{\rm{lift}} &= \bm{\psi}({\sf X}')
:=[\bm{\psi}(\bm{x}^{(1)}_1), \bm{\psi}(\bm{x}^{(2)}_1), \ldots, \bm{\psi}(\bm{x}^{(K)}_1)] \\
\end{aligned}
\right\}.
\]

Next, we explain the setting of MPC. 
The cost function $J$ is set so that we minimize in a fixed horizon $N_{\rm p}$ the error between the temperature of the output water (in the third layer of tank) and the reference temperature of $40\,\si{\degreeCelsius}$, and the power input to the heat pump. 
The cost function is
\begin{equation}
\label{costJ}
{\it{J}} = \sum_{i=0}^{N_{\mathrm{p}}-1} \left \{ Q\left  ( y_{i+1} - 40 \right )^2 + R u_i^2 \right \},
\end{equation}
where $Q$ and $R$ are positive weights. 
For the following simulations, the time step of the optimization is set at $h=60\,\si{s}$ to prevent chattering, and
the horizon $N_{\rm{p}}$ is set at $10$ (that is, $10h=600\,\si{s}$). 
The two weights are set as $Q=1$ and $R=10$, implying that we intend to keep the power input as low as possible.
The final time of closed-loop simulation is set as $1200\,\si{s}$, which is as long enough time from the first specification of the continuity of warm-water supply. 
This is relevant in practical situations.

\subsection{Numerical experiments}

We show two numerical experiments of the control of the warm-water supply system using BluSTL.  
The experiments were performed with MATLAB (R2020a). 
The initial value of the state variable $\bm{x}$ is set to $15\,\si{\degreeCelsius}$ for each component. 

\begin{figure}[t]
\centering
\includegraphics[width=0.46\textwidth]{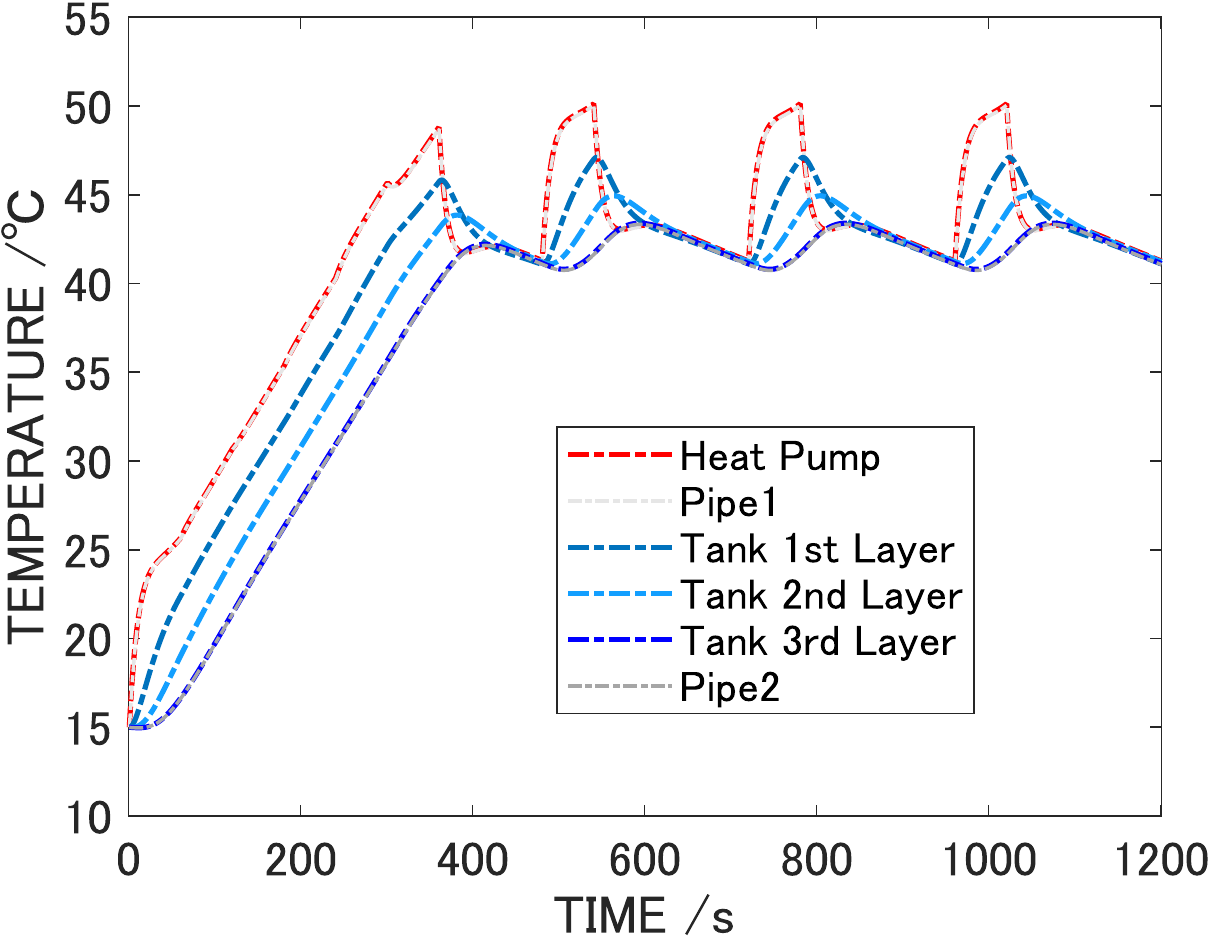} 
\setlength\abovecaptionskip{3pt}
\caption{Simulation of Koopman-MPC with STL specifications: water temperature of each component (state)}
\label{f:Ktemp}
\end{figure}

\begin{figure}[t]
\centering
\includegraphics[width=0.46\textwidth]{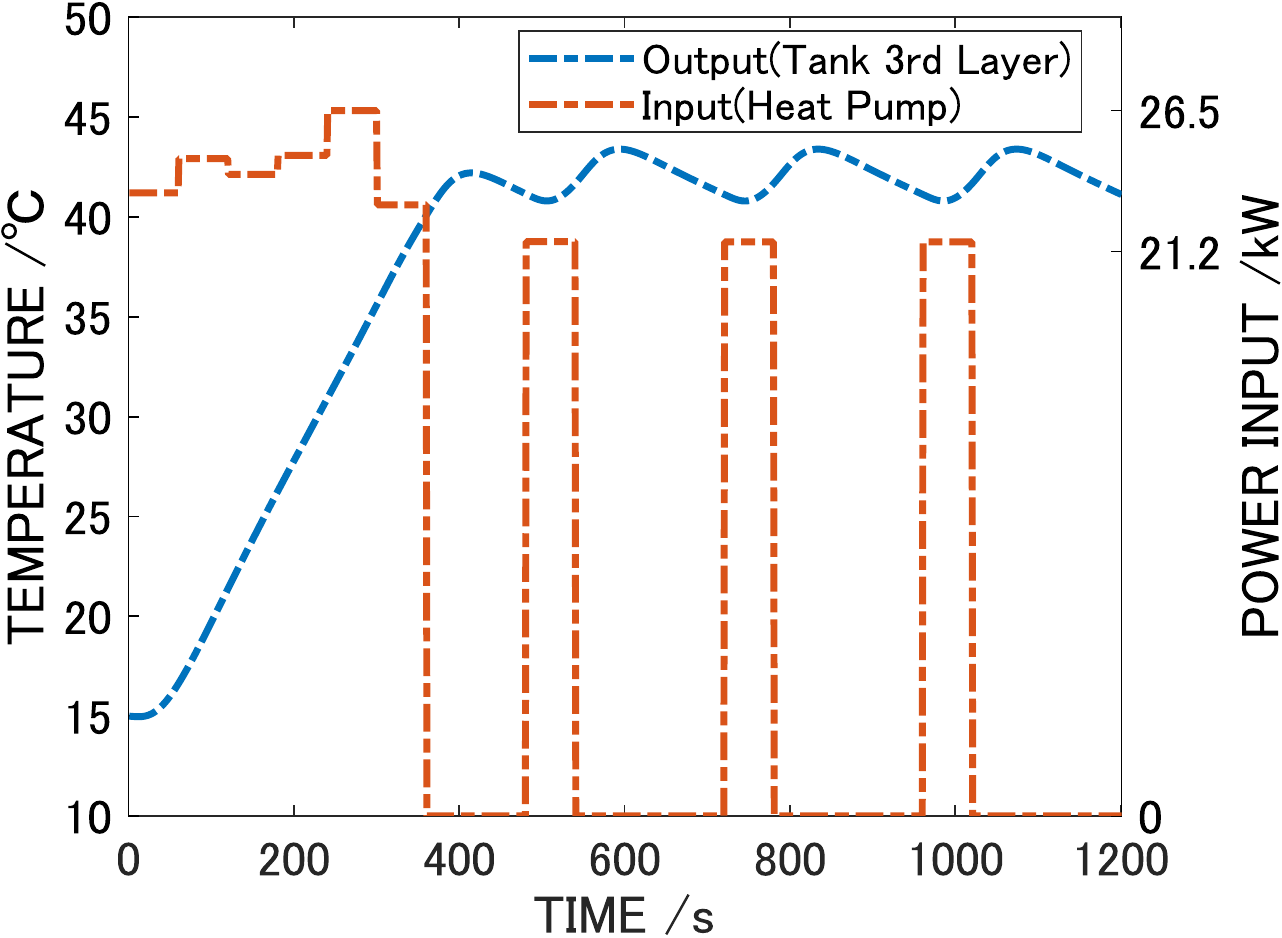} 
\setlength\abovecaptionskip{3pt}
\caption{Simulation of Koopman-MPC with STL specifications: water temperature of the third layer of tank (output) and the power input to heat pump (input) \vspace{0.7cm}}
\label{f:OUTIN}
\end{figure}

First, we show the time response of the warm-water supply system using the Koopman-MPC with STL specifications in Figs.\,\ref{f:Ktemp} and \ref{f:OUTIN}. 
Fig.\,\ref{f:Ktemp} shows the change of water temperature in each component of Fig.\,\ref{f:loop}. 
Fig.\,\ref{f:OUTIN} shows the output and input of the system. 
In Fig.\,\ref{f:Ktemp}, the water temperature increases in the order of pipe 1, first to third layer of the tank, and pipe 2. 
This is relevant from the configuration of the system and results from the primary delay and heat loss in the tank and pipes.
In Fig.\,\ref{f:OUTIN}, the output, namely, the water temperature of the third layer of the tank, is kept above $40\,\si{\degreeCelsius}$ from $420\,\si{s}$ to the end time while satisfying Eq.\,\eqref{stltank3}. 
In addition, the heat pump satisfies the specification of Eq.\,\eqref{stlhp}: 
its power input is kept above $80\%$ of the rated value and under the rated value during the operation, and is also kept at almost $0\,\si{kW}$ under no operation (that is, while the heat pump is stopped). 
Here, the output is fluctuated in the range between $40\,\si{\degreeCelsius}$ and $45\,\si{\degreeCelsius}$. 
This is due to the piecewise constant property of input under the zeroth-order hold with $h$. 


Second, we perform the feasibility test of execution of the Koopman-MPC with STL when the initial states and the start value $420\,\si{s}$ in the STL specification of Eq.\,\eqref{stltank3} are changed as in Table~\ref{tb:Range}.

\begin{table}[t]
  \begin{center}
    \caption{Feasibility test of Koopman-MPC with STL specifications}
    \label{tb:Range}
	\begin{tabular}{ccccccc} \toprule[0.5mm]
	\backslashbox{initial}{start} & 240 & 300 & 360 & 420 & 480 & 540 \\ \hline
	5				 & 0 & 0 & 0 & 0 &0 & 1 \\ \hline
	10				& 0 & 0 & 0 & 0 & 1 & 1 \\ \hline
	15				& 0 & 0 & 0 & 1 & 1 & 1 \\ \hline
	20				& 0 & 0 & 1 & 1 & 1 & 1 \\ \hline
	25				& 0 & 1 & 1 & 1 & 1 & 1 \\ \hline
	30				& 1 & 1 & 1 & 1 & 1 & 1 \\ \hline
	35				& 1 & 1 & 1 & 1 & 1 & 1 \\ \bottomrule[0.5mm]
	\end{tabular}
\end{center}
0:infeasible \\
1:feasible
\end{table}
The initial states are increased by $5\,\si{\degreeCelsius}$ from $5\,\si{\degreeCelsius}$ to $40\,\si{\degreeCelsius}$,  latter of which is the target value of the output. 
In addition, the start value $420\,\si{s}$ of Eq.\,\eqref{stltank3} is decreased by time step $60\,\si{s}$ from $600\,\si{s}$ of the horizon. 
Table~\ref{tb:Range} indicates that the feasible range is enlarged as the initial states approach the target value. 
This is mainly due to how long it takes for the water temperature to rise. 
In this sense, the feasibility range is highly dependent on the performance of the heat pump.

\subsection{Comparison of linear and Koopman MPCs}
\label{s:4.3}

\begin{figure}[t]
\centering
\includegraphics[width=0.46\textwidth]{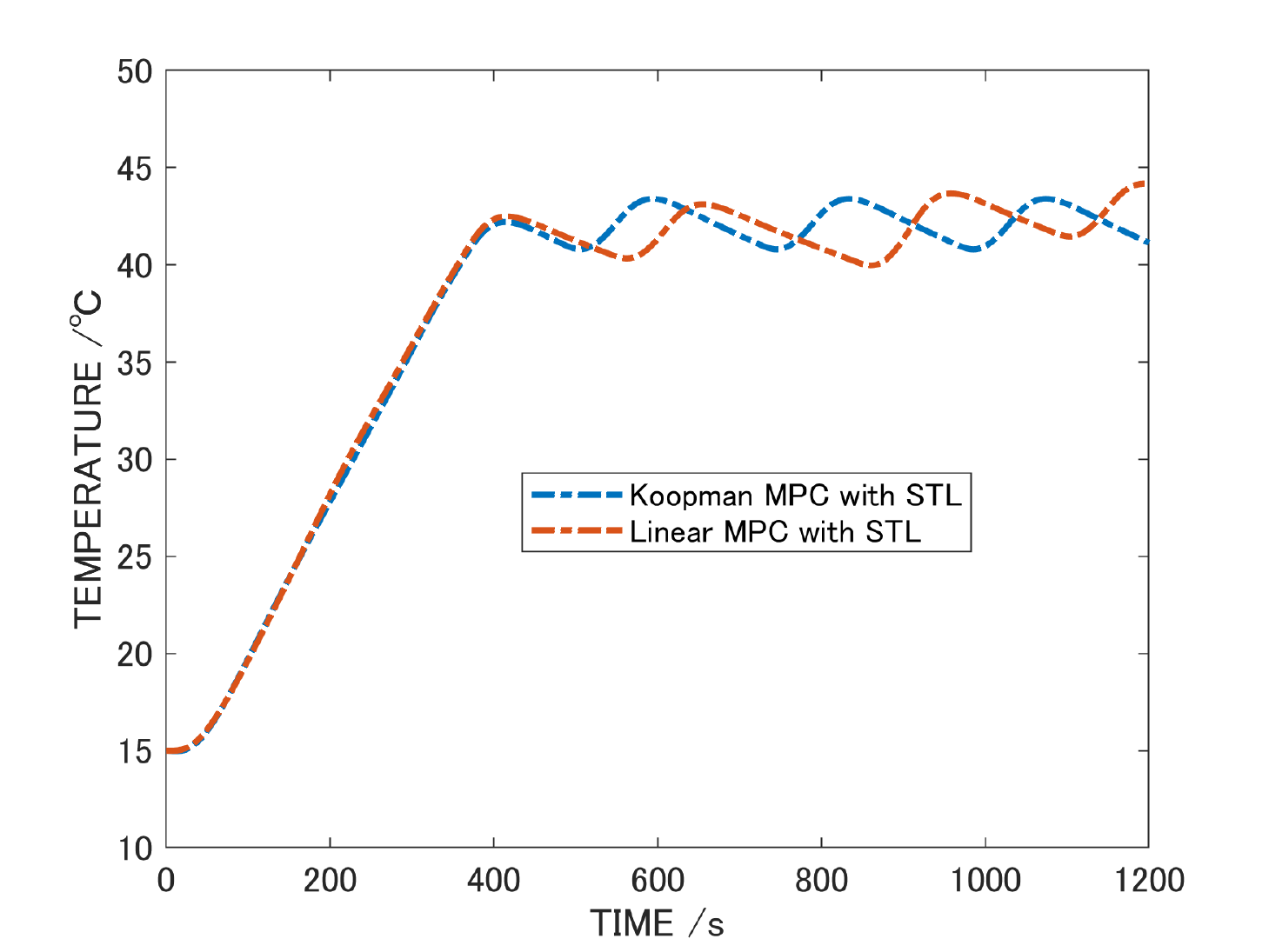} 
\caption{Comparison of Koopman and linear MPCs with STL specifications: output}
\label{f:com1}
\end{figure}
\begin{figure}[t]
\centering
\includegraphics[width=0.46\textwidth]{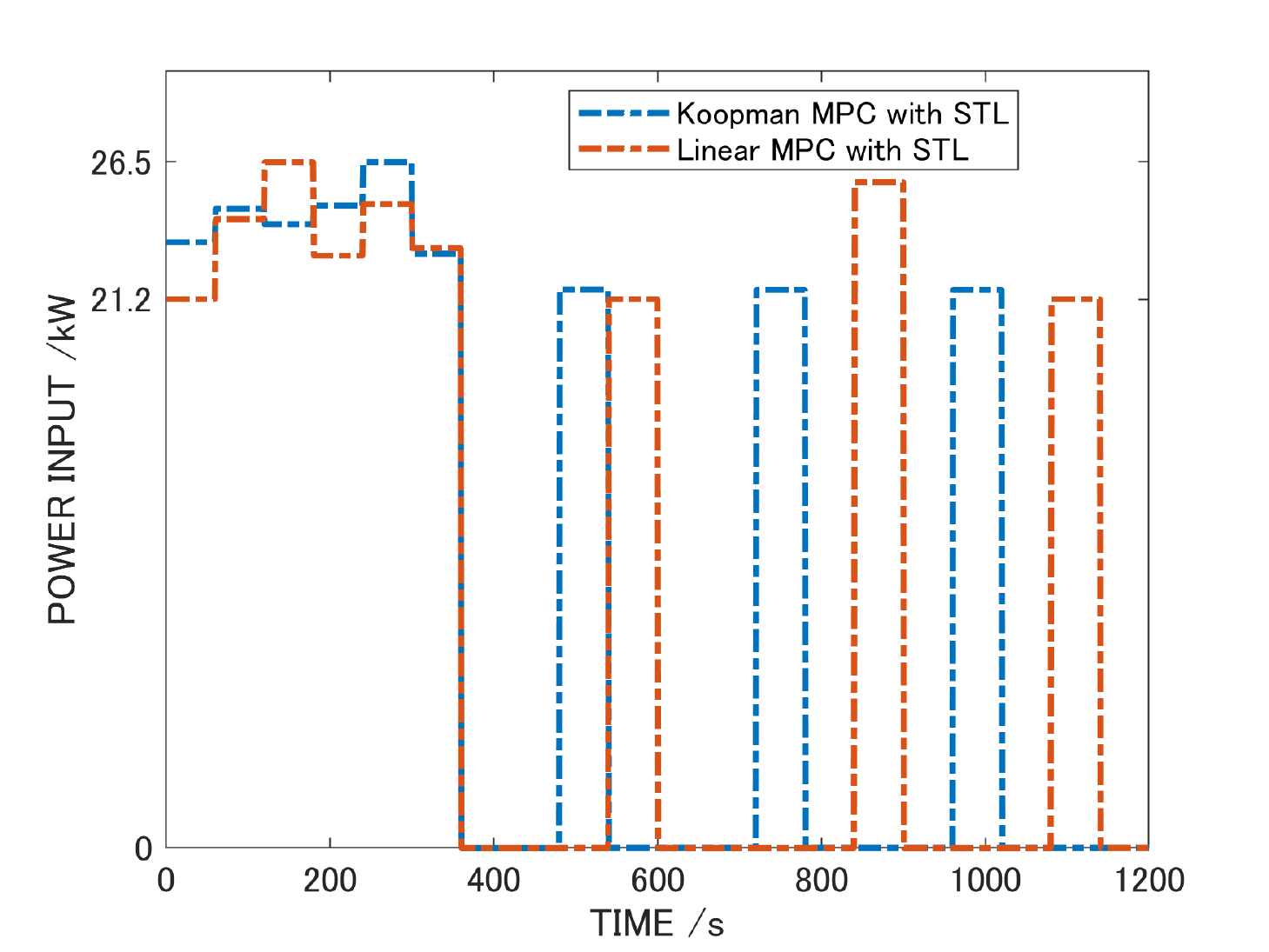}
\caption{Comparison of Koopman and linear MPCs with STL specifications: input}
\label{f:com2}
\end{figure}

We compare the control responses of the Koopman-MPC with STL specifications (K-MPC) and the linear MPC with STL specifications (L-MPC) based on local linearization.
Comparisons of the output and input are shown in Figs.\,\ref{f:com1} and \ref{f:com2}.
The time responses of temperature with the K-MPC and L-MPC are close in Fig.\,\ref{f:com1}.
A difference is found after the transient lasting until about 400\,s, where the small oscillations around 40$\rm {}^\circ C$ are delayed from each other.
The delay is clearly shown in the input of Fig.\,\ref{f:com2}, where the two time series behave in an anti-phase manner. 
This is because the slight difference in model accuracy used in the L-MPC and K-MPC exists in the simulations.

\section{Conclusions}
\label{s:5}
In this paper, we formulated the Koopman-MPC with STL specifications and applied it to the control problem of the warm-water supply system. 
This paper showed that the Koopman-MPC with STL specification guarantees a wider set of feasible initial states of the safe supply operation by comparison with the linear MPC with STL specifications.

Future directions are summarized as follows.
Without assuming that the outside temperature is constant, we will incorporate time-series prediction of disturbance into the model estimation. 
Since the performance of the linear predictor depends on the choice of the observables $\bm{\psi}$, we examine to improve the control performance by selecting a different set of observables. 
In addition, we examine another STL specification such as suppression of output temperature fluctuation from the viewpoint of dialysis treatment.

\section*{Comment to revision}
The authors claimed the numerical results presented in the first version of this preprint and \cite{MYST:SICE22}, that is, Figures~4 and 5, were not correct. 
Precisely, the results with the linear MPC with STL specifications were not correct. 
The authors revised these figures and sentences according to the correct results.


\renewcommand{\thetable}{\Alph{section}\arabic{table}}
\renewcommand{\theequation}{\Alph{section}\arabic{equation}}

\appendix

\section{Details of nonlinear dynamic model}
The nonlinear dynamic model of the warm-water supply system considered in this paper is presented below:
\[
\left.
\label{nonlinear_keisuu}
\begin{array}{cl}
\dot{x}_1  =& a_{1}x_1 +a_{2}x_6+ a_{3}u\\\vspace{0.5mm}
\dot{x}_2  =& a_{4}x_1 + a_{5}x_2 + a_{6}w\\
\vspace{0.5mm}
\dot{x}_3  =& a_{7}x_2 + a_{8}x_3 + a_{9}x_4 + a_{10}x_3^2 + a_{11}x_4^2 + \\\vspace{0.5mm}
& a_{12}x_3^2x_4 + a_{13}x_3x_4^2 + a_{14}x_3^3 +a_{15}x_4^3 + a_{16}w \\\vspace{0.5mm}
\dot{x}_4  =& a_{17}x_3 + a_{18}x_4 + a_{19}x_5 + a_{20}x_3^2 + a_{21}x_4^2 + \\\vspace{0.5mm}
& a_{22}x_5^2 + a_{23}x_3^2x_4 + a_{24}x_3x_4^2 + a_{25} x_4^2x_5 + \\\vspace{0.5mm}
&a_{26}x_4x_5^2 +a_{27}x_3^3 + a_{28}x_4^3 + a_{29}x_5^3 + a_{30}w\\\vspace{0.5mm}
\dot{x}_5  =& a_{31}x_4 + a_{32}x_5 +a_{33}x_4^2 + a_{34}x_5^2+a_{35}x_4^2x_5 + \\\vspace{0.5mm}
& a_{36}x_4x_5^2 + a_{37}x_4^3 + a_{38}x_5^3 +a_{39}w \\\vspace{0.5mm}
\dot{x}_6  =& a_{40}x_5 +a_{41}x_6 + a_{42}w \\
\end{array}
\right\}
\]
The coefficients of the model are summarized in Table~\ref{tb:keisuu}.
\begin{table}[t]
  \begin{center}
    \caption{Coefficients of the nonlinear dynamic model 
    }
    \label{tb:keisuu}
{
    \begin{tabular}{cccccc}  \toprule[0.5mm]
    $a_{1}$   &  $-9.8\times10^{-2}$ &      $a_{15}$   &  $3.0\times10^{-6}$          &  $a_{29}$      &      $3.0\times10^{-6}$   \\ \hline
     $a_{2}$   &  $4.0\times10^{-2}$  &      $a_{16}$  & $3.0\times10^3$            &   $a_{30}$     &     $2.0\times10^3$ \\   \hline            
     $a_{3}$   &  $9.8\times10^{-2} $ &      $a_{17}$  &      $  1.1  $                    &   $a_{31}$    & $1.1$   \\  \hline  
     $a_{4}$   & $3.8$ 	                 &    $a_{18} $   &    $ -2.0\times 10^3 $     & $a_{32}$      & $-3.0\times10^3$\\   \hline
    $a_{5}$    & $-2.4\times10^{2} $   &  $a_{19} $  &  $1.1  $                           & $a_{33}$    &$1.7\times10^{-3}$ \\\hline
    $a_{6}$   &  $2.4\times10^{2} $    &   $a_{20} $   &  $ 1.7\times 10^{-3} $       &  $a_{34}$  &$-1.7\times10^{-3}$ \\\hline
    $a_{7}$   & $3.0\times10^{-2} $   &   $a_{21} $   &  $-3.4\times 10^{-3} $    &  $a_{35}$   & $6.0\times10^{-6}$ \\\hline
    $a_{8}$  & $-3.0\times10^{3} $    &   $a_{22} $   &  $ 1.7\times 10^{-3} $          &  $a_{36} $  &$- 6.0\times10^{-6}$ \\\hline
    $a_{9}$  & $1.1 $                     &   $a_{23} $   &  $6.0\times 10^{-6} $         &  $a_{37} $ & $3.0\times10^{-6}$ \\\hline
    $a_{10}$  & $-1.7\times10^{-3} $  &   $a_{24} $    &  $ -6.0\times 10^{-6} $     &  $a_{38} $ & $-3.0\times10^{-6} $ \\\hline
    $a_{11}$  & $1.7\times10^{-3} $  &    $a_{25} $    & $ -6.0\times 10^{-6} $     & $a_{39} $ & $3.0\times10^{3}$ \\\hline
    $a_{12}$  & $-6.0\times10^{-6} $  &   $a_{26} $   & $ 6.0\times 10^{-6} $     & $a_{40} $ & $3.8$ \\\hline
    $a_{13}$  & $6.0\times10^{-6} $    &   $a_{27}$     &     $3.0\times10^{-6} $     & $a_{41}$ & $-2.4\times10^{2}$ \\\hline
    $a_{14}$    &  $-3.0\times10^{-6}$  &   $a_{28}$     &      $-6.0\times10^{-6}$  & $a_{42} $ & $2.4\times10^{2}$ \\  \bottomrule[0.5mm]
    \end{tabular}
}
  \end{center}
\end{table}

\end{document}